\PassOptionsToPackage{switch}{lineno}
\documentclass[conference]{IEEEtran}

\usepackage{microtype}

\IEEEoverridecommandlockouts
\usepackage{cite}
\usepackage{amsmath,amssymb,amsfonts,mathtools}
\usepackage{graphicx}
\usepackage{textcomp}
\usepackage[table,dvipsnames]{xcolor}
\usepackage{colortbl}
\def\BibTeX{{\rm B\kern-.05em{\sc i\kern-.025em b}\kern-.08em
    T\kern-.1667em\lower.7ex\hbox{E}\kern-.125emX}}

\usepackage{booktabs}   

\usepackage[linesnumbered,ruled,vlined]{algorithm2e}
\usepackage{setspace}
\usepackage{algpseudocode}
\usepackage{multicol}

\usepackage{circledsteps}
\pgfkeys{/csteps/inner color=RedViolet}
\pgfkeys{/csteps/outer color=ForestGreen}
\pgfkeys{/csteps/fill color=YellowGreen}

\SetKw{kwand}{and}

\usepackage{longtable}
\usepackage{xcolor}
\usepackage[binary-units=true]{siunitx}
\usepackage{url}

\usepackage{array}      
\usepackage{multirow}
\usepackage{bigstrut}    
\usepackage{adjustbox}
\usepackage{hhline}
\usepackage{mleftright}

\usepackage{graphicx}
\usepackage{tikz}
\usetikzlibrary{arrows,shapes}
\usetikzlibrary{arrows.meta}
\usetikzlibrary{positioning}
\usetikzlibrary{calc} 
\usetikzlibrary{decorations.pathreplacing} 
\usetikzlibrary{patterns}
\usetikzlibrary{backgrounds}
\usepackage{marvosym}   
\usetikzlibrary{decorations.pathmorphing} 

\tikzset{>=latex} 

\usepackage[hashEnumerators,smartEllipses,fencedCode]{markdown}

\usepackage[symbol]{footmisc}

\newcommand{\equivcls}[2]{%
  #1/\!{\sim}_#2%
}

\usepackage{listings}

\IEEEtriggeratref{41}

\usepackage{lineno}

\def\tabfontsize{\footnotesize}

\usepackage[colorinlistoftodos,color=yellow!20,]{todonotes} 

\usepackage{hyperref}
\hypersetup{colorlinks,
            breaklinks,         
	    linkcolor={blue!80!black},
            citecolor={red!70!black},
            urlcolor={blue!70!black}
           }

\lstdefinelanguage[RISC-V]{Assembler}
{
  alsoletter={.}, 
  alsodigit={0x}, 
  morekeywords=[1]{ 
    lb, lh, lw, lbu, lhu,
    sb, sh, sw,
    sll, slli, srl, srli, sra, srai,
    add, addi, sub, lui, auipc,
    mv, li,
    xor, xori, or, ori, and, andi,
    slt, slti, sltu, sltiu,
    beq, bne, bnez, beqz, blt, bge, bltu, bgeu,
    j, jr, jal, jalr, ret,
    scall, break, nop
  },
  morekeywords=[2]{ 
    .align, .ascii, .asciiz, .byte, .data, .double, .extern,
    .float, .globl, .half, .kdata, .ktext, .set, .space, .text, .word
  },
  morekeywords=[3]{ 
    zero, ra, sp, gp, tp, s0, fp,
    t0, t1, t2, t3, t4, t5, t6,
    s1, s2, s3, s4, s5, s6, s7, s8, s9, s10, s11,
    a0, a1, a2, a3, a4, a5, a6, a7,
    ft0, ft1, ft2, ft3, ft4, ft5, ft6, ft7,
    fs0, fs1, fs2, fs3, fs4, fs5, fs6, fs7, fs8, fs9, fs10, fs11,
    fa0, fa1, fa2, fa3, fa4, fa5, fa6, fa7
  },
  morecomment=[l]{;},   
  morecomment=[l]{\#},  
  morestring=[b]",      
  morestring=[b]'       
}

\def\zero{\ensuremath{0}}
\def\one{\ensuremath{1}}
\def\unknown{\ensuremath{{\mkern-2mu\times\mkern-2mu}}}

\newcommand{\psem}[2]{\sem\mleft((#1, #2)\mright)}
\newcommand{\ptrc}[2]{\trc\mleft((#1, #2)\mright)}
\DeclareMathOperator{\sem}{\mathit{s}}

\DeclarePairedDelimiter{\eqcl}{[}{]}
\def\trc{t}

\DeclareMathOperator{\abs}{\mathit{k}}
\DeclareMathOperator{\gpred}{pred}
\DeclareMathOperator{\gsucc}{succ}
\DeclareMathOperator{\gwrite}{write}
\DeclareMathOperator{\gread}{read}
\DeclareMathOperator{\gkill}{kill}
\DeclareMathOperator{\flowdef}{def}
\DeclareMathOperator{\flowuse}{use}
\DeclareMathOperator{\flowmeet}{\wedge}
\DeclareMathOperator{\flowop}{op}

\DeclareMathOperator{\flowand}{and}
\DeclareMathOperator{\flowmin}{min}
\DeclareMathOperator{\floweval}{eval}
\DeclarePairedDelimiter{\norm}{\lvert}{\rvert}

\def\captionsize{\footnotesize}

\newcommand\fig[1]{Fig.~{#1}}
\newcommand\tab[1]{Table~{#1}}
\newcommand\algo[1]{Algorithm~{#1}}

\newcommand\bb[1]{\texttt{\bfseries{#1}}}

\newcommand\var[1]{\texttt{#1}}

\newcommand{\ground}{%
    \def\cellwidth{2.4mm}%
    \def\cellheight{2.4mm}%
    \begin{tikzpicture}[baseline=-0.9mm,
       s/.style={rectangle, draw=gray, text=gray, line width=0.2mm,
                 text width = \cellwidth, minimum height = \cellheight,
	         align=center, inner sep=0pt, node distance=0mm},]%
    \node[s] () {};%
    \end{tikzpicture}%
}

\begin{document}
\urlstyle{rm}
%
%

\title{BEC:~Bit-Level Static Analysis for Reliability \\against Soft Errors}

\author{
\IEEEauthorblockN{Yousun Ko}
\IEEEauthorblockA{\textit{Department of Computer Science and Engineering} \\
\textit{Yonsei University}\\
Seoul, Republic of Korea \\
yousun.ko@yonsei.ac.kr} \\%
\and
\IEEEauthorblockN{Bernd Burgstaller}
\IEEEauthorblockA{\textit{Department of Computer Science and Engineering} \\
\textit{Yonsei University}\\
Seoul, Republic of Korea \\
bburg@yonsei.ac.kr} \\
}

\maketitle

\begin{abstract}
Soft errors are a type of transient digital signal corruption that occurs
in digital hardware components such as the internal flip-flops of CPU 
pipelines, the register file, memory cells, and even internal communication
buses. Soft errors are caused by environmental radioactivity, magnetic
interference, lasers, and temperature fluctuations, either unintentionally, or
as part of a deliberate attempt to compromise a system and expose confidential
data.

We propose a bit-level error coalescing~(BEC) static program analysis and its two use cases to
understand and improve program reliability against soft errors. The BEC 
analysis tracks each bit corruption in the register file and classifies
the effect of the corruption by its semantics at compile time. The usefulness
of the proposed analysis is demonstrated in two scenarios,
fault injection campaign pruning, and reliability-aware program transformation.
Experimental results show that bit-level analysis pruned up to \SI{30.04}{\percent}
of exhaustive fault injection campaigns (\SI{13.71}{\percent} on average), without loss
of accuracy. Program vulnerability was reduced by up to \SI{13.11}{\percent} (\SI{4.94}{\percent} on average)
through bit-level vulnerability-aware instruction scheduling. The analysis
has been implemented within LLVM and evaluated on the \makebox{RISC-V} architecture. 

To the best of our knowledge, the proposed BEC analysis is
the first bit-level compiler analysis for program reliability against soft
errors. The proposed method is generic and not limited to a specific computer
architecture.
\end{abstract}



\section{Introduction}\label{sec:intro}

Soft errors---also known as transient hardware faults---are one of the most
common threats to the reliable operation of digital devices. Soft errors
temporarily alter one or more bits in hardware, thereby corrupting the
execution semantics of a program. Soft errors can happen in any hardware
component that is exposed to lasers, radiation~\cite{May:1979,Baumann:2005},
magnetic interference~\cite{Baffreau:2002}, and temperature
fluctuations~\cite{Jagannathan:2012}, either accidentally or intentionally to
compromise a computer system.

Soft errors can be masked without any observable
effect on program execution (e.g., if after the occurrence of a soft error in a CPU register
the corrupted bit is overwritten by the application logic and thus reverted to a correct state).
However, soft errors that are not masked will propagate through the software stack
and may lead to
catastrophic system failure by corrupting sensitive data, or
altering the control flow of an application and thereby potentially granting unauthorized access to critical
program paths.
Mitigation of soft errors is thus an essential concern, particularly
for safety-critical systems such as avionics, space, autonomous vehicles,
nuclear reactor control systems, and life support devices.

Hardware-level soft error mitigation methods, such as event detection and correction~(EDAC), are effective but make hardware more expensive to design, manufacture, and operate. For instance, EDAC has been reported to increase the device cost by \SI{40}{\percent} and to increase power consumption~\cite{EDAC:2020}. 
More importantly, the protection level of hardware methods is limited. 
For instance, EDAC can detect up to double-bit flips and correct single-bit flips,
which makes software-level countermeasures indispensable~\cite{Schroeder:2011}.
A recent report cites radiati on effects as the
cause for 2128~single-bit upsets in the SRAM of a satellite during a
286~day low-orbit mission~\cite{Noeldeke:2021}. The SRAM was equipped with EDAC
circuitry.
Fault attacks~\cite{Yuce:2018} are an outstanding example of
soft-error exploitation, to compromise a system by injecting faults to
divert the control flow of a program and thereby expose security-critical information being
processed by the program. 

In this work, we propose a bit-level error coalescing~(BEC) analysis, to understand and improve
program reliability against soft errors at bit-level. The proposed BEC analysis tracks each
bit corruption in the register file and classifies the effect by its semantics
at compile-time. Thus the results of the analysis can be
utilized by other compiler analyses and optimizations.

BEC analyzes each bit of a register value separately and independently.
This complies better with the nature of soft errors
and provides more accurate analysis results compared to value-level analyses.
For a soft error to propagate, the target register must
contain a live value (i.e., a value that will be read in the future). But
not every bit of a live value may be live. For instance, a bit-shifting operation will
shift out some bits of a register while preserving other bits (albeit in different
bit positions). 
Furthermore, BEC identifies which live bit corruptions are equivalent 
in their effects. Let us assume that a corrupted
bit value is relocated to a new bit location via a shift operation without
affecting any other operation in between. Then, the effect of a bit
corruption that occurred before the shift operation will be equivalent to
the effect of a bit corruption that occurred at the new bit location after the shift
operation. 

We have implemented the proposed BEC analysis within LLVM~16.0.0 and demonstrated
its effectiveness for eight distinctive benchmarks on the \makebox{RISC-V}~\cite{RISCVISA}
architecture. Experimental results show that the proposed bit-level analysis
prunes up to \SI{30.04}{\percent} of the fault injection campaigns (\SI{13.71}{\percent}
on average) compared to value-level analysis,
and reduces the program vulnerability by up to \SI{13.11}{\percent} through bit-level vulnerability-aware instruction scheduling (\SI{4.94}{\percent} on average).

This paper makes the following contributions:
\begin{enumerate}
    \item We propose BEC, the first bit-level static program analysis that exploits the semantics of operations to track and classify the effect of bit corruptions due to soft errors.
    \item We propose an abstract bit-value analysis, which extends the scope of the analysis from sequences of instructions to global scope.
    \item We validate the BEC analysis empirically and show its soundness. 
    \item We demonstrate the effectiveness of the BEC analysis with two use cases, fault injection campaign pruning and vulnerability-aware instruction scheduling, on eight distinctive benchmarks.
    \item We have implemented the BEC analysis within LLVM and made it available to the public~\cite{BEC:github}.
\end{enumerate}

The remainder of this paper is organized as follows.
We introduce the relevant background in Section~\ref{sec:prelim}.
Section~\ref{sec:motivating_example}
illustrates the proposed bit-level analysis with a motivating example, followed
by the formal description in Section~\ref{sec:bit_level_analysis}. Section~\ref{sec:validation} empirically
validates the proposed method, before introducing the two use cases of the
BEC analysis in Section~\ref{sec:usecase:fi_pruning} and Section~\ref{sec:usecase:vul_anal}.
We discuss the related work in Section~\ref{sec:related_work} and draw our conclusions in Section~\ref{sec:conclusion}.

\section{Background and Notation\label{sec:prelim}}
A program $P = \{p_0, \ldots, p_{n-1}\}$ consists of $n=\vert P\vert$ program
points (i.e., instructions). Each program point maintains the same set
$V=\{v_0, \ldots, v_{m-1}\}$ of data points (i.e., variables). The number of
data points of the system that executes program~$P$ is denoted by $m=\vert
V\vert$. In the case of physical hardware, $m$ indicates the number of
registers provided by the underlying hardware architecture, and $m$ is
conceptually infinite if no underlying hardware is specified (e.g., with the virtual registers of
LLVM). Data points may refer to memory cells if data in memory is modeled by a
compiler. Data point $v_y = [v_y^{w-1}, \ldots, v_y^{0}]$ is of
bit-width~$w$. We use the little-endian notation for the bit representations of
data points, thus, $v_y^i$ indicates the value of the $i$-th least-significant bit~(LSB)
of data point~$v_y$. For the sake of simplicity, we assume that 
all data points are of the same bit-width.

$P$ provides the scope of the temporal structural locations of a program and $V$ the scope of the spatial locations of the underlying hardware where soft errors may occur. We denote the Cartesian product of the two orthogonal scopes, $F=P\times V$, as the overall \emph{fault space}.
A single \emph{fault site} on a bit $v_y^i$ at a program point $p_x$ is denoted by $(p_x, v_y^i) \in F$.
We define $F^0$ to be an empty tuple $()$.

A control-flow graph (CFG, \cite{Dragonbook:2006}) is a directed graph
$\langle P,E,e,x\rangle$, where nodes are program points $p\in P$, and edges
$E\subseteq P\times P$ represent a transfer of control between program points.
The unique entry and exit nodes are denoted by~$e$ and~$x$, respectively. For
program point~$p$ and CFG~$G$, $\gpred(p)$ and $\gsucc(p)$ are the sets of
predecessor and successor program points. For program point $p$,
$\gwrite(p)$, $\gread(p)$, and $\gkill(p)$ are the sets of data points written,
read, and killed at $p$, respectively. Data points in $\gread(p)$ must be live
before program point $p$, and data points in $\gkill(p)$ are no longer live
after program point $p$.

We use a variation of definition-use chains~\cite{ACDI} to connect the data
points in a CFG. By $\flowdef(p,v)$ we denote the set of program points~$p'$
that define data point~$v$ and there is a CFG path from~$p'$ to~$p$ that does
not re-define (kill) data point~$v$.
Similarly, by $\flowuse(p,v)$ we
denote the set of program points~$p'$ that use data point~$v$ from~$p$ and
there is a path from~$p$ to~$p'$ that does not re-define (kill) data
point~$v$. 
Note that $\flowuse(p,v)$ is sensitive to control-flow dependencies. 
$\norm{\flowdef(p,v)}$ can be greater than~1, which means that 
data flow is not limited to static single assignment~(SSA) form~\cite{SSA}.

As the proposed BEC analysis is a bit-level analysis, we introduce a notion for the known bit values of data points at arbitrary program points.
$\abs(p, v)$ denotes the bit values of data point~$v$
after program point~$p$. Likewise, $\abs(p, v^i)$ denotes the bit value
of the $i$-th bit from the LSB of data point~$v$ after program point~$p$.
$\abs(p,v^i)$ is an abstract interpretation~\cite{Cousot:1977} of a bit value, which is \zero\ or \one\
if the bit is known to be zero or one in any of the temporal
states of the fault site $(p, v^i)$, $\top$ if it is not possible to
determine the actual value of the corresponding bit at compile-time, and $\bot$
if the value is undefined. The abstraction function is defined as
$\gamma(\zero) = \{\zero\}$, $\gamma(\one) = \{\one\}$, $\gamma(\top)=\{\zero,
\one\}$, $\gamma(\bot)=\emptyset$, and the concretization function is given as
$\alpha(\{\zero\})=\zero$, $\alpha(\{\one\})=\one$, $\alpha(\{\zero,
\one\})=\top$, and $\alpha(\emptyset)=\bot$. The concept of $\abs(p, v^i)$ is
comparable to \emph{KnownBits} in LLVM and \emph{tnum} in the Berkeley Packet
Filter~(BPF, \cite{DBLP:conf/usenix/McCanneJ93}).

\section{Motivating Example}\label{sec:motivating_example}

\begin{figure}
\centering
\setlength{\tabcolsep}{1.4mm}
\begin{tabular}{l}
\input{tikz/ex_mot/src.tex}
\end{tabular}
\caption{Motivating example to count the number of years that are even but not a multiple of four, inspired by the concept of leap year.}\label{fig:mot:src}
\end{figure}

\begin{figure*}
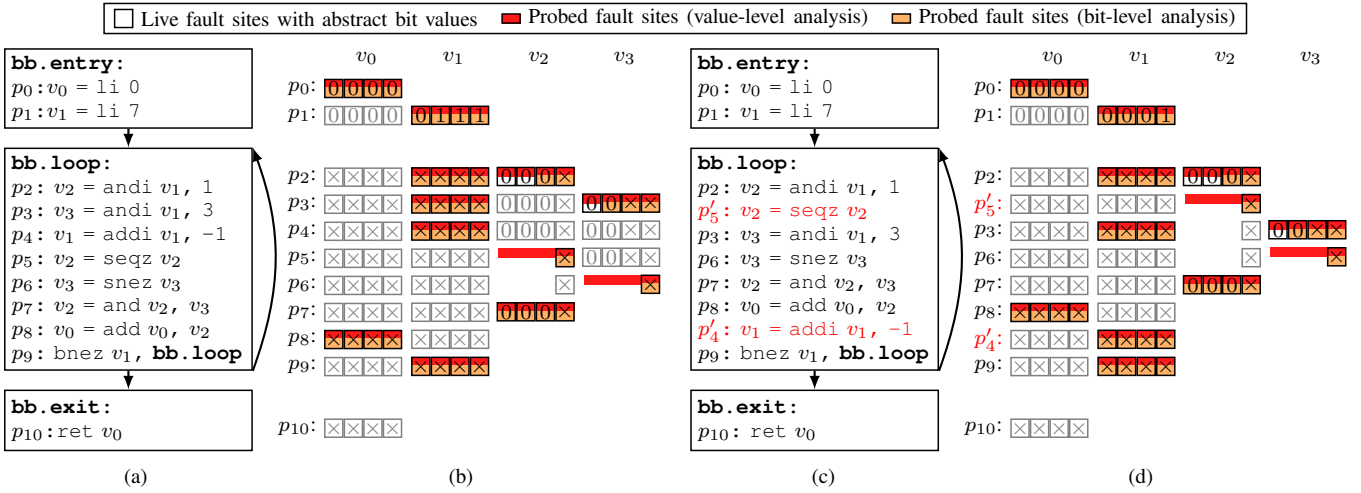

\setlength{\tabcolsep}{0mm}%
\begin{tabular}{ccc}%
\multicolumn{3}{c}{%
\def\cellwidth{2.4mm}%
\def\cellheight{2.4mm}%
\def\fiheight{1.2mm}%
\begin{tikzpicture}[%
  bb/.style={rectangle, draw=black, line width=0.2mm, text width = 2.5cm, minimum height=0.6cm, align=left, node distance=2.5mm},
  fi/.style={rectangle, draw=black, line width=0.2mm, text width = \cellwidth, minimum height = \cellheight, align=center, inner sep=0pt, node distance=0mm},
  fiv/.style={rectangle, fill=red!90, line width=0.2mm, text width = \cellwidth, minimum height = \fiheight, align=center, inner sep=0pt, node distance=0mm},                                          
  fib/.style={rectangle, fill=orange!60, line width=0.2mm, text width = \cellwidth, minimum height = \fiheight, align=center, inner sep=0pt, node distance=0mm},
]
\footnotesize

\node[fi](wb){}; 
\node[right = 0mm of wb](wbt){Live fault sites with abstract bit values}; 
\node[fiv, draw=black, right = 2mm of wbt](rb){}; 
\node[right = 0mm of rb](rbt){Probed fault sites (value-level analysis)};
\node[fib, draw=black, right = 2mm of rbt](ob){};
\node[right = 0mm of ob](obt){Probed fault sites (bit-level analysis)};
\node[bb, right = -4mm of wb, yshift=0mm, text width=150mm, minimum height=4mm] (legend) {};
\end{tikzpicture}%
}\\%
\input{tikz/ex_mot/fig.tex}%
&\rule{4mm}{0mm}&%
\input{tikz/ex_sched/fig.tex}%
\\%
\multicolumn{3}{l}{
\captionsize
\hspace{16mm}(a)\hspace{40mm}(b)\hspace{45mm}(c)\hspace{39mm}(d)
}\\
\end{tabular}
\caption{(a)~CFG and (b)~fault sites of the motivating example from \fig{\ref{fig:mot:src}}. 
With fault sites, the x-axis presents data points~(variables).
The y-axis refers to program points~(instructions), which are labeled by their corresponding instructions
from the CFG (labels~$p_0$--$p_{10}$).
Live fault sites are data and program points that contain live values, depicted
by white boxes with known bit values. Boxes are colored if the fault sites are
identified as subjects for vulnerability tests by value-level analysis~(red) or
bit-level analysis~(orange). The right half of the figure~(c,d) depicts the
motivating example after instruction rescheduling to minimize the live fault
sites in bits. Note that the number of instructions to be executed and the
number of fault injection runs required remain unchanged after bit-level
vulnerability-aware instruction scheduling, at a reduction of the number of live fault
sites by \SI{15.4}{\percent}.
}\label{fig:mot:fi_map}
\end{figure*}

Our motivating example counts the number of years that are even but not a
multiple of four. It was inspired by the concept of leap year but simplified
for the sake of exposition. For the same reason we confine the discussion of
our motivating example to a 4-bit architecture. \fig{\ref{fig:mot:src}}
depicts the source code of the program in~C. The \var{for} loop
(lines~\ref{line:leapyear:loop}--\ref{line:leapyear:cond}) iterates from~7 to~1
to accumulate in variable~\texttt{res} the number of years that meet the
before-mentioned condition.

\fig{\ref{fig:mot:fi_map}}a presents the CFG of our motivating
example. The proposed analysis anticipates the bit-level semantics (i.e.,
side effect) of each instruction of the \makebox{RISC-V} target architecture~\cite{RISCVISA}
represented in three-address code format~\cite{Dragonbook:2006}.

\fig{\ref{fig:mot:fi_map}}b illustrates the fault sites. Fault sites constitute the
program points where the soft errors that occurred in the underlying hardware are first observed
at the software level as bit flips. Soft errors can happen in any part of the hardware where machine status
is stored in bits, i.e., the register file, internal flip-flops of pipelines,
memory buses, memory cells etc., but soft errors must be visible to the software,
for instance, as a bit corruption of the register file, to divert the behavior of the
program. 
A bit flip of a fault site represents any soft error that happened on any
hardware component and ended up corrupting a bit at the fault site. Thus the
probability of observing a bit flip at a fault site due to soft errors may
differ by hardware design and operational environments, but the consequence of
a bit flip manifested at a particular fault site is analyzable in software. We use a single event
upset model per run as it is the dominant attack model for studies from both academia and
industry~\cite{Dixit:2009, Sridharan:2013, Schirmeier:2015}.
Bits flipped by soft errors remain on the system until overwritten.

In the presence of cycles in the CFG of a program, a single
fault site can be reached multiple times during a program run, and the faults that occur across different visitations of a fault site are semantically distinguishable.
For example, faults that occur at data point~$v_1$ at program point~$p_2$ in basic block~\bb{bb.loop} in different loop iterations are semantically different and may lead to different outcomes. For
the sake of simplicity, we use four-bit unsigned values for the spatial
fault sites of our running example in \fig{\ref{fig:mot:fi_map}}.
The LSB of a fault site is its rightmost bit. 

The BEC analysis first computes the prospective bit values, $\abs(p_x, v_y^i)$ for all $(p_x, v_y^i)\in F$. We use $\unknown$ instead of $\top$ in the motivating examples to denote that a bit value is unknown at compile-time.
Suppose we have $\abs(p, v)=\zero\zero\unknown\one$, then the value of data point~$v$ at program point~$p$ can be either \zero\zero\zero\one\ or \zero\zero\one\one.
Throughout this paper, we use boxes~(\ground) to depict the abstract bit values of fault sites. For instance, in \fig{\ref{fig:mot:fi_map}}a at program point~$p_1$, data point~$v_1$ is initialized to constant~7. Thus, $\abs(p_1, v_1)=\zero\one\one\one$, as shown in the second column of \fig{\ref{fig:mot:fi_map}}b. 
Inside the loop body~(program points $p_2$--$p_9$ of basic block~\bb{bb.loop}), the bits of
data point~$v_1$ are unknown because data point~$v_1$ is an
induction variable and the analysis information must hold for all possible
temporal states of the loop.

Based on the analyzed bit values of fault sites, the BEC analysis identifies and classifies the effect of bit
corruptions at compile-time. The static classification of bit corruptions at
all possible fault sites in a program is useful for understanding the 
vulnerability of a program and enhancing its reliability against soft
errors without additional run-time overhead. In the remainder of this section, we will
outline the two use cases we have explored to demonstrate the usefulness
of the proposed bit-level analysis.

\subsection{Use Case~1: Fault Injection Pruning}
A fault injection campaign is an effective technique to assess the
reliability and robustness of a program or a system. It 
involves the systematic injection of faults into a running program to
evaluate its behavior under faulty conditions. To determine the impact of a bit
corruption at a given fault site, the running program is suspended at the clock cycle
prior to the fault site, the bit under investigation is flipped,
and the program is resumed until the outcome becomes
certain. 
The outcome of the program after the fault injection is compared with the
golden output (i.e., the trace of events and the output obtained from a
fault-free execution of a program).

The probability of a program malfunctioning in the presence of soft errors
can be obtained by conducting fault injection runs on each 
spatial fault site of each temporal fault site---for each bit in
the register file of a processor at each cycle---and
counting the number of malfunctioning
cases. While such an exhaustive fault injection campaign can provide valuable insights into a
program's reliability, it is an extremely resource-intensive process. The large
number of faults required to obtain a statistically meaningful figure for
reliability renders exhaustive fault injection infeasible for large and
complex programs. A principled method for the identification
and selection of a representative subset of faults to inject without
sacrificing analysis precision is therefore required 
to accelerate fault injection campaigns. 

The BEC analysis can reduce the number of fault
injection sites without any loss of coverage or accuracy. In other words, the
analysis results of an exhaustive fault injection campaign can be achieved by performing
only a subset of the fault injection runs. As discussed in the following, the
colored boxes in \fig{\ref{fig:mot:fi_map}}b illustrate how the proposed
BEC analysis can accelerate a fault injection campaign.

Inject-on-read~\cite{Berrojo:2002,Schirmeier:2015,Sangchoolie:2015} is a method proposed to accelerate fault
injection campaigns on hardware at gate-level, by injecting faults only when the
fault site is read. For instance, data point~$v_0$ in \fig{\ref{fig:mot:fi_map}} is a return value that
accumulates the number of the years of interest, thus, it is live throughout
the lifespan of the function. But fault injection runs on data point~$v_0$
are required only at program point~$p_8$ per loop iteration, and the
results of the fault injection runs at program point~$p_8$ are identical to
the fault injection runs performed at any program point after the
previous program point that accessed the variable, which is program
point~$p_8$ of the previous loop iteration. The inject-on-read method is
efficient, yet its analysis is performed on {\em values}. In \fig{\ref{fig:mot:fi_map}}b
we marked the fault sites in red for which fault injection is required
by the inject-on-read method.

The proposed bit-level analysis can further identify which bits {\em within\/}
a value require a fault injection run. For example, data point~$v_2$ keeps the
result of instruction~$p_2$ and $p_5$ in \fig{\ref{fig:mot:fi_map}}b,
which encodes the condition \texttt{year\%2 == 0} in
line~\ref{line:leapyear:cond} of \fig{\ref{fig:mot:src}}. After the execution
of instruction~$p_2$, which masks out all but the LSB of
data point~$v_2$, it holds that $\abs(p_2, v_2)=\zero\zero\zero\unknown$. Knowing that
instruction~{\tt seqz} at program point~$p_5$ tests $v_2$ for zero,
we note that any corruption that flips any one of the bits $v_2^1, v_2^2$, or $v_2^3$ from~0 to~1 will result in the same
negative test result. 
Consequently, only one fault injection is
required among the bits $v_2^1, v_2^2$, and $v_2^3$ at program point~$p_2$. As indicated by the orange color for
data point~$v_2$ at program point~$p_2$ in
\fig{\ref{fig:mot:fi_map}}b, our bit-level analysis injects only bit~$v_2^1$,
whereas the prior approach injects all three bits. Because
program point~$p_2$ is part of the loop, the savings achieved by our
analysis pertain to each temporal fault site (i.e., one per loop iteration) of
program point~$p_2$. Unlike our running example, the registers of
contemporary processors are 32- or even 64-bit wide. Capitalizing on
those two facts, our bit-level analysis yields noticeable gains (as discussed
in Section~\ref{sec:usecase:fi_pruning}).

Not only can the proposed bit-level analysis classify which fault sites share
an identical response to soft errors, it can further analyze which fault
sites are dead, meaning that bit corruption on such fault sites are
known to be ineffective. Fault sites $(p_5, v_2^1), (p_5, v_2^2)$, and $(p_5, v_2^3)$ are dead in \fig{\ref{fig:mot:fi_map}}b because a corruption of any of those 
bits will be masked by the \texttt{and} operation at
program point~$p_7$.

The fault sites required to be probed by fault injection runs by the proposed
BEC analysis are marked by orange boxes in \fig{\ref{fig:mot:fi_map}}b, in
contrast to the red boxes which are fault sites required to be probed by
value-level analysis. With the motivating example in \fig{\ref{fig:mot:fi_map}}
in conjunction
with the given loop bounds, the number of fault injection runs required by
value-level analysis is $288$\footnote[2]{$4+4+7\times(4+4\times 4+3\times4+2\times4)= 288$}, whereas the number of
fault injection runs required by the proposed bit-level analysis is
$225$\footnote[3]{$4+4+7\times(4+4\times4+2+1+4+3+1) = 255$}. Thus, the number of fault injection runs
saved is $1-\frac{225}{288}=\SI{21.8}{\percent}$.

\subsection{Use Case~2: Bit-level Vulnerability-aware Instruction Scheduling}

The size of the spatial and temporal fault surface for soft errors is one of
the main metrics to determine a program's vulnerability. If Program~A
completes a task in the same number of cycles as Program~B, but requires more hardware
resources, then Program~A is more vulnerable to soft errors because of its
larger spatial fault surface than Program~B. Similarly, if Program~A
takes longer than Program~B to complete the same task with the same amount of
resources, then program~A is more susceptible to soft errors regarding its
temporal fault surface. Thus, it is important to minimize the number of
instructions and hardware resources required to complete a task to improve the
overall reliability of a program.

The fault surface of a program can be determined by the number of live fault
sites in data for every program point executed by a program run. In
\fig{\ref{fig:mot:fi_map}}b, the number of live fault sites for a program run 
is 681\footnote[8]{$3\times 4+7\times(8 \times 4 + 8\times 4 + 4\times 4 + 2 \times 1 + 3 \times 4 +1) + 4 = 681$}. As in the case of data point~$v_2$ at program
points~$p_5$ and $p_6$, and data point~$v_3$ at program
point~$p_6$ in \fig{\ref{fig:mot:fi_map}}b,
dead fault sites identified by the proposed bit-level analysis
can be exploited to reduce the overall vulnerability of the program by
rescheduling of instructions. 

In the example, data points~$v_0$ and $v_1$ are live at
almost every program point. As a result, rescheduling
instructions would not decrease the number of fault sites for $v_0$ or
$v_1$ on any program point. However, data points~$v_2$ and $v_3$ are
temporary within each iteration and only carry one fault site per program
point after program points~$p_5$ and $p_6$ in \fig{\ref{fig:mot:fi_map}}a
have been executed. Thus, there is room for reducing fault sites
by instruction scheduling on these two variables. 
\fig{\ref{fig:mot:fi_map}}c shows the modified sequence of
instructions and \fig{\ref{fig:mot:fi_map}}d shows its fault sites of the same motivating example
from~\fig{\ref{fig:mot:src}} after instruction rescheduling to minimize the live
fault sites. 

The number of fault sites per program point is reduced from four to one after
program point~$p_5'$ for data point~$v_2$ and program point~$p_6$ for
data point~$v_3$ in \fig{\ref{fig:mot:fi_map}}c. Thus, these instructions are executed as early as
possible in \fig{\ref{fig:mot:fi_map}}c. Temporary variables are retired as early
as possible for the same purpose. With the new sequence of instructions, the
total number of fault sites of the program is reduced by \SI{15.4}{\percent}
($=1-\frac{576}{681}$).
Note that the number of instructions to be executed and the number of fault injection runs required remain unchanged after instruction rescheduling.

\section{Bit-level Analysis for Reliability}\label{sec:bit_level_analysis}


To classify the effect of soft errors at fault sites, 
we introduce the notion of \emph{fault index},
denoted by $\psem{p_x}{v_y^i}$. A fault index labels the effect induced by a soft
error at fault site $(p_x, v_y^i) \in F$.
We use $s_0$ to denote the intact execution of the program, and $s_0=\sem\mleft(F^0\mright)=\sem\mleft(()\mright)$.
The set of fault indices~$S$ is defined as $S=\{\psem{p_x}{v_y^i} \mid (p_x, v_y^i)\in F\}\cup \{\sem_0\}$.
For a given program, if bit flips at two distinct fault
sites $(p, u^i)$ and $(q, v^j)$ have
the same effect on program execution, we consider
these fault sites as {\em equivalent}.
We employ equivalence relations~\cite{Paulson:equiv:classes} to represent the
equivalence of fault indices based on the equivalence of their associated fault
sites in terms of the underlying program semantics.

An equivalence relation $R = \equivcls{S}{R}$ constitutes the set of all equivalence classes over~$S$
induced by a binary relation~$\sim_R$. We write $\sem_x \sim_R \sem_y$ for $\sem_x, \sem_y
\in S$, denoting that the fault indices $\sem_x$ and $\sem_y$ are in the same 
equivalence class under~$R$. Hence, the effects of faults
at their associated fault sites are equivalent according to the above definition. 

The equivalence class of $\sem_x$ under
$\sim_R$ is defined as $[\sem_x]_R = \{\sem_y\in S \mid \sem_x \sim_R \sem_y\}.$
For instance, if $\eqcl*{\sem_x}_R=\{\sem_x, \sem_y\}$, it holds that $[\sem_x]_R = [\sem_y]_R$. 
We define the merge of two equivalence classes as $R\mleft[\eqcl*{s_x}_R \cup \eqcl*{s_y}_R\mright] = \mleft((R \setminus [s_x]_R)\setminus[s_y]_R\mright) \cup \{\eqcl*{s_x}_R \cup \eqcl*{s_y}_R\}$ for any $s_x,s_y\in S$. By abuse of notation, we merge arbitrary equivalence classes as $R\eqcl*{X} = R\eqcl*{\bigcup_{x\in X}[x]_R}$ for any $X \subset S$.

The proposed BEC analysis computes a safe approximation of the equivalence relation~$\equivcls{S}{R}$ over the
fault indices~$S$ of a program. If the analysis determines that
$[\psem{p}{v^i}]_R = [\psem{q}{u^j}]_R$, then the effects of soft
errors at fault sites $(p, v^i)$ and $(q, u^j)$ are known to be equivalent.
Conversely, if the analysis fails to establish the equivalence of two fault sites
wrt.~soft errors, they may still be equivalent but cannot be shown so within the 
analysis approximations (this is further discussed in Section~\ref{sec:validation}).
BEC employs the abstract bit values of a fault site and is thus conducted
in two steps: (1)~a global abstract bit-value
analysis~(Section~\ref{subsec:bit_val_anal}), which is a forward data-flow
analysis at the granularity of the individual bits of data points, 
and (2)~an analysis to coalesce fault
indices~(Section~\ref{subsec:collapsing_anal}), which is a backward 
data-flow analysis that identifies and classifies the corruption of the program semantics due to soft
errors based on bit values.

\subsection{Global Abstract Bit Value Analysis}\label{subsec:bit_val_anal}

\begin{figure}
\centering
\begin{tabular}{ccc}
\begin{tabular}{c}
\begin{tikzpicture}
  \tikzset{>={Latex[width=1.2mm,length=1.4mm]}}
  \footnotesize
  
  \node[circle] (top) at (0, 1.2) {$\top$};
  \node[circle] (0) at (-0.5, 0.6) {\zero};
  \node[circle] (1) at (0.5, 0.6) {\one};
  \node[circle] (bot) at (0, 0) {$\bot$};

  \draw[-] (top) -- (0);
  \draw[-] (top) -- (1);
  \draw[-] (0) -- (bot);
  \draw[-] (1) -- (bot);
\end{tikzpicture}
\end{tabular}
&
%
%
{%
\setlength{\tabcolsep}{.7mm}
\footnotesize
\begin{tabular}{c|cccc}
$\flowmeet$&$\bot$&$0$&$1$&$\top$\\
\hline
$\bot$&$\bot$&$0$&$1$&$\top$\\
$0$&$0$&$0$&$\top$&$\top$\\
$1$&$1$&$\top$&$1$&$\top$\\
$\top$&$\top$&$\top$&$\top$&$\top$\\
\end{tabular}
}
&
%
%
{%
\setlength{\tabcolsep}{.7mm}  
\footnotesize
\begin{tabular}{c|cccc}
$\flowand$&$\bot$&$0$&$1$&$\top$\\
\hline
$\bot$&$\bot$&$\bot$&$\bot$&$\top$\\     
$0$&$\bot$&$0$&$0$&$0$\\   
$1$&$\bot$&$0$&$1$&$\top$\\   
$\top$&$\top$&$0$&$\top$&$\top$\\
\end{tabular}
}\\[-1mm]
{\footnotesize (a)}&{\footnotesize (b)}&{\footnotesize (c})\\[-2mm]
\end{tabular}
\caption{Bit-level analysis: (a)~lattice representation of bit values, (b)~meet operator, and (c)~bit-wise $\flowand$ operator.\label{fig:knownbits}}
\end{figure}
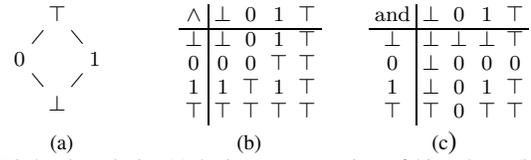

To analyze the effects of soft errors at the granularity of individual
bits, we first identify the bit values of all data points
i.e., $\abs(p_x, v_y^i),\forall (p_x, v_y^i) \in F$,
by performing a forward data-flow analysis across the entire program.
Our analysis is inspired by a value-level constant propagation
algorithm by Wegman and Zadek~(SC, \cite{Wegman:const:prop}), which we extend
to bit-level analysis. 
We deliberately locate our analysis at a late stage
within the pass sequence of the LLVM backend to benefit from target-specific strength reduction optimizations
that lower arithmetic operations to bit-level operations and thereby
increase the opportunity for the application of our analysis. 
SSA form is already deconstructed at this stage, and our analysis computes $\flowuse(p_x, v_y^i), \forall (p_x, v_y^i) \in F$ as defined in Section~\ref{sec:prelim}.
LLVM provides a bit-level analysis named \emph{KnownBits} for straight-line sequences of
code. We extend this analysis to
support inter-basic-block (global) data flow.

\fig{\ref{fig:knownbits}}a depicts the lattice representation of bit values
employed with the analysis: $\bot$ (undefined) is used for a bit at a
particular program point that has not seen an assignment yet; $0$ (or $1$) is
used for a bit of a data point where it is certain that the bit-value will be
zero (or one) on all paths to the respective program point that have been
considered so far, and $\top$ (unknown and overdefined) denotes a bit value that cannot be
determined at compile-time. For instance, a bit is unknown and overdefined if the bit value is zero on some paths (or loop iterations) and one on others.
For each bit, the computed information can only raise in the lattice.

\begin{algorithm}[t]
\caption{Bit-Value Analysis}\label{algo:bit_anal}
\SetKwInput{KwInput}{Input}
\SetKwInput{KwOutput}{Output}
\DontPrintSemicolon

\KwInput{program point $p \in P$ and data point $v \in V$}
\KwOutput{$\abs(p, v)$, the abstract bit-value of $v$ after the execution of $p$}

\ForAll{$u \in \gread(p)$\label{algo:bit:read_begin}} {
  \ForAll{$o \in \flowdef(p,u)$\label{algo:bit:merge_begin}} {
    \For{$i\leftarrow  \{ 0$, \ldots, $w-1$\}\label{algo:bit:funnel_begin}}{
     $\abs(p, u^i) \gets \abs(p, u^i) \flowmeet \abs(o, v^i)$\label{algo:bit:forward}
   }
  }
}
\ForAll{$v \in \gwrite(p)$} {\label{algo:bit:def} 
  \For{$i\leftarrow \{ 0$, \ldots, $w-1$\}}{\label{algo:bit:write:start}
   $\abs(p, v^i) \gets \flowop_p\left(\{\abs(p, u) \mid u \in \gread(p)\}\right)$ \label{algo:bit:abs_op}
  }
}

\end{algorithm}

Algorithm~\ref{algo:bit_anal} describes the analysis for
a given program point~$p$ and data point~$v$. Because of join points in the underlying
CFG, multiple definitions may reach~$p$ for any of the data points
it reads. The purpose of the loop starting at line~\ref{algo:bit:merge_begin}
is to merge all definitions that reach data point~$u\in \gread(p)$.
The innermost loop iterates over
all the bits of data point~$u$. It uses the meet operator~($\flowmeet$)
from \fig{\ref{fig:knownbits}}b. For instance, the meet of definitions zero and
one sets the corresponding bit to overdefined, i.e.,~$\flowmeet(0,1)=\top$.

In case program point~$p$ computes values (e.g., the bit-wise~$\flowand$
of two values), lines~\ref{algo:bit:write:start}--\ref{algo:bit:abs_op} in Algorithm~\ref{algo:bit_anal}
iterate over all bits in the corresponding data point~$v$ to perform the
computation ($\flowop_p$ in line~\ref{algo:bit:abs_op}) 
of the program point in the abstract domain of the lattice. 
As an example, \fig{\ref{fig:knownbits}}c provides the definition of the bit-wise~$\flowand$
operation. The definition of an operation ensures that the abstract value of a
bit can only move upward in the lattice. 
We note that the analysis permits any number of values computed at a program point (line~\ref{algo:bit:def} in Algorithm~\ref{algo:bit_anal}), which is a generalization of the three-address code used in the motivating example in Section~\ref{sec:motivating_example}.


%
%
%
\subsection{Bit-level Fault Index Coalescing Analysis}\label{subsec:collapsing_anal}
The fault index coalescing analysis is a backward data-flow analysis that (1)~identifies which fault sites mask soft errors and (2)~classifies which fault sites result in equivalent program semantics when corrupted by
a bit flip.

\begin{algorithm}[t]
\caption{Fault Index Coalescing Analysis}\label{algo:coalescing}
\SetKwInput{KwInput}{Input}
\SetKwInput{KwOutput}{Output}
\SetKw{Continue}{continue}
\DontPrintSemicolon

\KwInput{a program $P=\{p_0, \ldots, p_{n-1}\}$, a set of data points $V=\{v_0, \ldots v_{m-1}\}$, and bit-width $w$}
\KwOutput{A set of equivalence classes $R = \equivcls{S}{R}$}
 \tcc{Initialization}
 $R \gets \mleft\{\{\sem_0\}\mright\}$ 

 \ForAll{$(p, v^i) \in F$\label{algo:coll:init_for}}
 {
   \If{$v \in \gwrite(p) \vee v \in \gread(p)$} {
     \If{$v \in \gkill(p)$} {
       $\eqcl*{\sem_0}_R \gets \eqcl*{\sem_0}_R \cup \mleft\{\psem{p}{v^i}\mright\}$\label{algo:coll:dead}
     } \Else {
       $R \gets R \cup \mleft\{\{\psem{p}{v^i}\}\mright\}$ \label{algo:coll:init}
     }
   }
 }
 \tcc{Iterative coalescing}
 \While{$R$ is updated\label{algo:coll:iter}}
 {
   \tcc{Intra-instruction coalescing}
   \ForAll{$p \in P$}
   {
     $R'_p \gets \mbox{intra\_instr\_coalescing}(p,V, w, R)$\label{algo:coll:intra}
   }
   
   \tcc{Inter-instruction coalescing}
   \ForAll{$(p,v^i) \in F$\label{algo:coll:inter}}
   {
     $R \gets R\eqcl*{\eqcl*{\psem{p}{v^i}}_R \cup \bigcap_{q\in \flowuse(p,v)}{\eqcl*{\psem{q}{v^i}}_{R'_q}}}$\rule{0mm}{4mm}\label{algo:coll:agreed}
   }
 }

\end{algorithm}

Algorithm~\ref{algo:coalescing} describes the fault
index coalescing algorithm, which assigns fault
indices $\sem\in S$ the same equivalence class if the effects of faults
occurring at those indices are identical (equivalent) according to the
semantics of the underlying program.
The algorithm returns an equivalence relation $R = \equivcls{S}{R}$.

The loop from line~\ref{algo:coll:init_for} in 
\algo{\ref{algo:coalescing}} initializes equivalence relation~$R$ for each data point~$v$ that is accessed (i.e., read or written) at program point~$p$. If the accessed data point~$v$ is not live after program point $p$, any faults occurring at $v$ after $p$ will be overwritten and masked. 
Soft errors at masked fault sites do not alter the program semantics.  
Thus masked fault sites are assigned to the equivalence class $s_0$~(line~\ref{algo:coll:dead} in \algo{\ref{algo:coalescing}}). For fault site $(p, v^i)\in F$, if data point~$v$ is accessed at and live after program point~$p$,
a new equivalence class is assigned to the fault site~(line~\ref{algo:coll:init} in \algo{\ref{algo:coalescing}}). 
Note that data points that are not accessed at a program point may be live and susceptible to soft errors, but the equivalence relation considers those only that are
accessed at the program point.
This is because the semantics of a program can only
be changed by reading a corrupted value regardless of when the corruption
happened. Thus we postulate that the effect of any faults that occurred at a data
point are the same until the program reaches the program point that reads the
data point.

After initialization, equivalence relation~$R$ is a set of singletons of all fault sites of~$F$.
It is interpreted as no live fault sites are identified to be equivalent nor
masked.
This is sound but not necessarily precise. The analysis results are
subsequently refined by coalescing equivalence classes within 
instructions~(intra-instruction coalescing, line~\ref{algo:coll:intra} in
\algo{\ref{algo:coalescing}}) and across instructions~(inter-instruction
coalescing, line~\ref{algo:coll:agreed} in \algo{\ref{algo:coalescing}}), in an
iterative manner.

\begin{algorithm}
\setstretch{1.07}
\caption{Intra-instruction Coalescing}\label{algo:intra_instr_coalescing}
 \SetKwInput{KwInput}{Input}
\SetKwInput{KwOutput}{Output}
\SetKw{Continue}{continue}
\DontPrintSemicolon

\KwInput{a program point $p\in P$, a set of data points $V=\{v_0, \ldots v_{m-1}\}$, bit-width $w$, and a set of equivalence classes $R = \equivcls{S}{R}$}
\KwOutput{A set of equivalence classes $R' = \equivcls{S}{{R'}}$}

   $R' \gets \emptyset$, $I \gets \{0, \ldots, w-1\}$
   
   \ForAll{$i \in I$}
   { 
     \If{$p: z = \mathtt{mv}~x$ \kwand $x,z \in V$}
     {\label{algo:intra:mv}
       $R' \gets R\eqcl*{\eqcl{\psem{p}{x^i}}_R \cup \eqcl{\psem{p}{z^i}}_R}$ 
     }
     \ElseIf{$p: z = \mathtt{xor}~x, y$ \kwand $x,y,z \in V$}
     {\label{algo:intra:xor:if}
       $R' \gets R\eqcl*{\eqcl{\psem{p}{x^i}}_R \cup \eqcl{\psem{p}{z^i}}_R}$\\[1mm]\label{algo:intra:xor:merge:start}
       $R' \gets R\eqcl*{\eqcl{\psem{p}{y^i}}_R \cup \eqcl{\psem{p}{z^i}}_R}$
       \label{algo:intra:xor:merge:end}
     }
     \ElseIf{$p: z = \mathtt{or}~x, y$ \kwand $x,y,z \in V$}
     {
       \If{$\abs(p, y^i) = 0$}
       {
         $R' \gets R\eqcl*{\eqcl{\psem{p}{x^i}}_R \cup \eqcl{\psem{p}{z^i}}_R} $
       }
       \ElseIf{$\abs(p, y^i) = 1$}
       {
         $R' \gets R\eqcl*{\eqcl{\psem{p}{x^i}}_R \cup \eqcl{\sem_0}_{R}} $\\
       }
        \ElseIf{$\abs(p, x^i) = 0$}
       {
         $R' \gets R\eqcl*{\eqcl{\psem{p}{y^i}}_R \cup \eqcl{\psem{p}{z^i}}_R} $
       }
       \ElseIf{$\abs(p, x^i) = 1$}
       {
         $R' \gets R\eqcl*{\eqcl{\psem{p}{y^i}}_R \cup \eqcl{\sem_0}_{R}} $
       }
     }
     \ElseIf{$p: z = \mathtt{and}~x, y$ \kwand $x,y,z \in V$}
     {
       \If{$\abs(p, y^i) = 0$}
       {\label{algo:intra:and:if:mirror:start}
         $R' \gets R\eqcl*{\eqcl{\psem{p}{x^i}}_R \cup \eqcl{\sem_0}_{R}} $
       }
       \ElseIf{$\abs(p, y^i) = 1$}
       {
         $R' \gets R\eqcl*{\eqcl{\psem{p}{x^i}}_R \cup \eqcl{\psem{p}{z^i}}_R} $ \label{algo:intra:and:if:mirror:stop}
       }
        \ElseIf{$\abs(p, x^i) = 0$}
       {\label{algo:intra:and:if}
         $R' \gets R\eqcl*{\eqcl{\psem{p}{y^i}}_R \cup \eqcl{\sem_0}_{R}} $ \label{algo:intra:and:merge}
       }
       \ElseIf{$\abs(p, x^i) = 1$}
       {\label{algo:intra:and:if1}
         $R' \gets R\eqcl*{\eqcl{\psem{p}{y^i}}_R \cup \eqcl{\psem{p}{z^i}}_R} $ \label{algo:intra:and:merge1}
       }
     }
     \ElseIf{$p: z = \mathtt{shr}~x, y$ \kwand $x,y,z \in V$}
     {
       \If{$i - \flowmin(p, y) < 0$}
       {
         $R' \gets R\eqcl*{\eqcl{\psem{p}{x^i}}_R \cup \eqcl{\sem_0}_{R}}$
       }
       \ElseIf{$y$ is constant \kwand $i - y \geq 0$}
       {
         $R' \gets R\eqcl*{\eqcl{\psem{p}{x^i}}_R \cup \eqcl{\psem{p}{z^{i-y}}}_R} $
       }
     }
     \ElseIf{$p: z = \mathtt{shl}~x, y$ \kwand $x,y,z \in V$}
     {
       \If{$i + \flowmin(p, y) \geq w$}
       {
         $R' \gets R\eqcl*{\eqcl{\psem{p}{x^i}}_R \cup \eqcl{\sem_0}_{R}} $
       }
       \ElseIf{$y$ is constant \kwand $i + y < w$}
       {
         $R' \gets R\eqcl*{\eqcl{\psem{p}{x^i}}_R \cup \eqcl{\psem{p}{z^{i+y}}}_R} $
       }
     }
   }

     \If{$p: \mathtt{[slt|beq|bne|bge|blt]}~x, y$ \kwand $x,y \in V$}
     {
       \ForAll{$i\in I, j \in \{0, \ldots, i-1\}, v \in \{x, y\}$} {
         \If{$\floweval(p, v^i) = \floweval(p, v^j)$}
         {
           $R' \gets R\eqcl*{\eqcl{\psem{p}{v^i}}_R \cup \eqcl{\psem{p}{v^j}}_R} $
         }
       }
   }

\end{algorithm}

Intra-instruction coalescing is depicted in
Algorithm~\ref{algo:intra_instr_coalescing} for a selection of \makebox{RISC-V}
instructions~\cite{RISCVISA}, but the proposed method is general and
applicable to any other instruction set architecture~(ISA).
During intra-instruction coalescing, equivalence classes are merged based on the semantics of the operation of a program point, applied to the abstract bit values of
the accessed data points.
For instance, for program point $p$ with bitwise operation $z = \mathtt{and}~x, y$ and
a bit of operand~$x$ known to be zero, i.e., $\abs(p, x^i) = 0$~(line~\ref{algo:intra:and:if} in
\algo{\ref{algo:intra_instr_coalescing}}), a soft error carried to
fault site $(p, y^i)$
is masked as a result of the \texttt{and} operation.
Thus, the equivalence class $[\psem{p}{y^i}]_R$ is merged with
$[s_0]_R$~(line~\ref{algo:intra:and:merge} in \algo{\ref{algo:intra_instr_coalescing}}). Conversely, if bit~$x^i$ of
operand~$x$ is known to be one (line~\ref{algo:intra:and:if1} in \algo{\ref{algo:intra_instr_coalescing}}), a fault on the
corresponding bit~$y^i$ will be propagated to the result bit~$z^i$. Thus, these
two equivalence classes are merged (line~\ref{algo:intra:and:merge1} in \algo{\ref{algo:intra_instr_coalescing}}).
Lines~\ref{algo:intra:and:if:mirror:start}--\ref{algo:intra:and:if:mirror:stop} in \algo{\ref{algo:intra_instr_coalescing}}
follow by commutativity of the $\mathtt{and}$ operation. 

A few instructions in our analysis are
oblivious to the abstract bit values of their operands. Instructions \texttt{mv} and \texttt{xor} belong to 
this category, and coalescing is conducted unconditionally. With
\texttt{xor} (line~\ref{algo:intra:xor:if} in \algo{\ref{algo:intra_instr_coalescing}}), any soft errors at $(p, x^i)$ or
$(p, y^i)$ are propagated to $(p, z^i)$ after the operation. Thus, the 
equivalence classes $[\psem{p}{x^i}]_R$ and $[\psem{p}{y^i}]_R$ are merged with
$[\psem{p}{z^i}]_R$~(lines~\ref{algo:intra:xor:merge:start} and
\ref{algo:intra:xor:merge:end} in \algo{\ref{algo:intra_instr_coalescing}}). 

For brevity, several utility functions have been introduced in
\algo{\ref{algo:intra_instr_coalescing}}. Function~$\flowmin(p, v)$ returns the minimum possible value of data point~$v$ at program point~$p$ considering
$\abs(p, v)$. Function~$\floweval(p, v^i)$ (partially) evaluates the instruction at 
program point~$p$ based on the abstract bit values of the operands, assuming
that a soft error occurred at fault site~$(p, v^i)$. If $p$ is a branch
instruction, the result of the evaluation is the target branch taken
by the instruction.

The temporary equivalence relation~$R'$ is introduced to defer the merge of the equivalence
classes until all fault sites have been visited during inter-instruction coalescing.
Equivalence class $[\psem{p}{v^i}]_{R}$ is merged and updated
only if $[\psem{q}{v^i}]_{R'_q}$ for all $q\in \flowuse(p,v^i)$ agree~(line~\ref{algo:coll:agreed} in
\algo{\ref{algo:coalescing}}) for a given $(p, v^i)\in F$. 

The iterative fault index coalescing analysis terminates when no further update of 
the equivalence classes in~$R$ occurs~(line~\ref{algo:coll:iter} in
\algo{\ref{algo:coalescing}}), thereby reaching a fixed
point~\cite{Dragonbook:2006}.
%
By Knaster-Tarski's fixed point theorem~\cite{Khedker:2009}, any monotone function on a complete lattice admits a least fixed point. 
The collection of all equivalence relations on a set forms a complete lattice~\cite{Ore:1942}, and the fault index coalescing analysis is monotonic:
it is performed backward along the dependency edges, from $\gwrite(p)$ to $\gread(p)$ and from $\flowuse(p, v)$ to $\flowdef(q, v)$. Thus, the iterative fault index coalescing algorithm is guaranteed to terminate.

\subsection{Coalescing Example}

\begin{figure*}
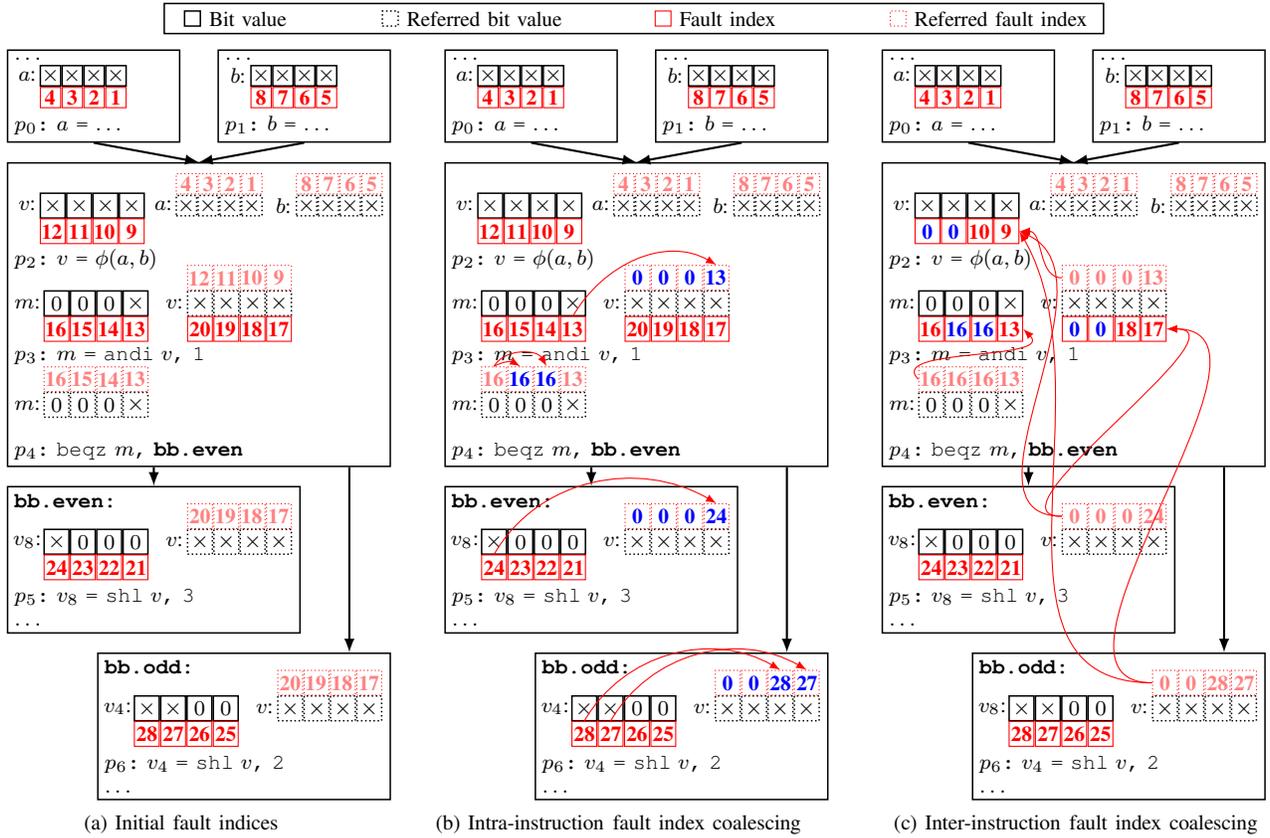

\centering
\setlength{\tabcolsep}{0mm}
\renewcommand{\arraystretch}{0}
\begin{tabular}{ccc}
\multicolumn{3}{c}{
\def\cellwidth{2.5mm}
\def\cellwidthsmall{2.1mm}
\def\rshifty{2mm}
\def\slmargin{1mm}
\def\sbmargin{0mm}
\def\sdistx{4mm}
\def\legenddistx{34mm}
\begin{tikzpicture}[
  tbox/.style={node distance=0mm, align=center},
  bb/.style={rectangle, draw=black, line width=0.2mm, text width = 2.5cm, minimum height=0.6cm, align=left, node distance=2.5mm},
  tag/.style={rectangle, draw=none, text width = \cellwidth+1mm, minimum height = 2mm, align=center, inner sep=0pt, node distance=0mm},
  s_def/.style={rectangle, draw=black, line width=0.2mm, text width = \cellwidth, minimum height = 2mm, align=center, inner sep=0pt, node distance=0mm},
  s_use/.style={rectangle, draw=black, densely dotted, line width=0.2mm, text width = \cellwidth, minimum height = 2mm, align=center, inner sep=0pt, node distance=0mm},
  r/.style={rectangle, draw=red, text=red, text width = \cellwidth, minimum height = 2mm, align=center, inner sep=0pt, node distance=0mm},
  shadow_r/.style={rectangle, draw=red, text=red!50, densely dotted, text width = \cellwidth, minimum height = 2mm, align=center, inner sep=0pt, node distance=0mm},
  s_def_small/.style={rectangle, draw=black, line width=0.2mm, text width = \cellwidthsmall, minimum height = 2mm, align=center, inner sep=0pt, node distance=0mm},
  s_use_small/.style={rectangle, draw=black, densely dotted, line width=0.2mm, text width = \cellwidthsmall, minimum height = 2mm, align=center, inner sep=0pt, node distance=0mm},
  r_small/.style={rectangle, draw=red, text=red, text width = \cellwidthsmall, minimum height = 2mm, align=center, inner sep=0pt, node distance=0mm},
  shadow_r_small/.style={rectangle, draw=red, text=red!50, densely dotted, text width = \cellwidthsmall, minimum height = 2mm, align=center, inner sep=0pt, node distance=0mm},
  descr/.style={rectangle, draw=none, text width = \legenddistx, minimum height = 2mm, align=left, inner sep=0pt, node distance=0mm},
  ]
  \footnotesize

  \node(zero) at (0,0) {};

  \node[bb, below = of zero, yshift=3mm, text width=123mm, minimum height=4mm] (legend) {};
  
  \node[s_def_small, left = of legend, xshift=5mm] (s_def) {};
  \node[descr, right = of s_def, xshift=1mm] (s_def_descr) {Bit value};
  
  \node[s_use_small, right = of s_def, xshift=\legenddistx-10mm] (s_use) {};
  \node[descr, right = of s_use, xshift=1mm] (s_use_descr) {Referred bit value};
  
  \node[r_small, right = of s_use, xshift=\legenddistx] (r_def) {};
  \node[descr, right = of r_def, xshift=1mm] (r_def_descr) {Fault index};
  
  \node[shadow_r_small, right = of r_def, xshift=\legenddistx-5mm] (r_use) {};
  \node[descr, right = of r_use, xshift=1mm] (r_use_descr) {Referred fault index};

\end{tikzpicture}
}\vspace{2mm}\\

\hspace{-4mm}
\input{tikz/anal_ex/init.tex}
&
\hspace{-4mm}
\input{tikz/anal_ex/local.tex}
&
\hspace{-4mm}
\input{tikz/anal_ex/global.tex}\vspace{2mm}\\

\captionsize
(a) Initial fault indices
& 
\captionsize
(b) Intra-instruction fault index coalescing 
&
\captionsize
(c) Inter-instruction fault index coalescing\\
\end{tabular}
\caption{Iterative fault index coalescing of a fork-after-join CFG snippet using 4-bit data points: (a)~initial fault indices are assigned to bits of data points, (b)~fault indices are coalesced within their instruction during the intra-instruction fault-index coalescing phase, and (c)~fault indices are coalesced across instructions during the inter-instruction fault index coalescing phase. Note that coalescing is a monotonic process that is performed backward along the dependency edges. The example code is in SSA form for brevity, but the proposed method is not limited to SSA form.%
}\label{fig:analysis_ex}
\end{figure*}

\fig{\ref{fig:analysis_ex}}a presents the initial fault indices of our coalescing example in red solid boxes. Fault indices are represented by integer identifiers such as $\psem{p_0}{a^0}=1$ and $\psem{p_5}{v_8^3}=24$. Fault index identifier~$0$ is reserved for the intact semantics, $s_0\in S$. The example uses data points of bit-width four, thus four red solid boxes are mapped for each data point.

Note that fault indices of data point~$v$ may differ before and after a
program point that reads the data point, even if the value of the data point
remains the same. For instance, in \fig{\ref{fig:analysis_ex}}a,
data point~$v$ written at program
point~$p_2$ is read at program point~$p_3$, and new fault indices are assigned
to data point~$v$ after the data point is live and will be read at program
points~$p_5$ and $p_6$. Thereby the analysis distinguishes a bit corruption at $v^j$ between
program points~$p_2$ and $p_3$ from the ones 
between program points~$p_3$ and $p_5$ or $p_6$. No new fault
indices are assigned to data point~$v$ at program points~$p_5$ and $p_6$ as
the data point is assumed to be killed at those program points.

The fault index coalescing analysis is an iterative process and is performed in two phases: (1)~intra-instruction fault index coalescing, where fault indices are coalesced within a program point~(\fig{\ref{fig:analysis_ex}b}), and (2)~inter-instruction fault index coalescing, where fault indices are coalesced across instructions~(\fig{\ref{fig:analysis_ex}c}). By repeating the two phases, fault index coalescing merges equivalence classes of fault indices backward along the dependency edges, until a fixed point is reached.


\fig{\ref{fig:analysis_ex}}b illustrates intra-instruction fault index
coalescing of our coalescing example. At program point~$p_3$, an \texttt{and}
operation is conducted on argument~$v$ and an immediate of bit
representation~\texttt{0001}. According to the intra-instruction fault
index coalescing rule of the \texttt{and} operation from 
\algo{\ref{algo:intra_instr_coalescing}}, soft errors at fault sites
$(p_3, v^1), (p_3, v^2),$ and $ (p_3, v^3)$ are masked, thus $\psem{p_3}{v^1} \sim_{R'} \psem{p_3}{v^2} \sim_{R'} \psem{p_3}{v^3} \sim_{R'} s_0$, where $R'$ is the temporary equivalence relation in \algo{\ref{algo:intra_instr_coalescing}}.

Intra-instruction fault index coalescing is performed among the data points read within a program point as well. For instance, in case of the instruction~\texttt{beqz \%mod, \bb{bb.even}} at program point~$p_4$, any soft errors occurring at bits where the value is known to be 0  will divert the control flow to \bb{bb.odd}. Thus $\psem{p_4}{m^1} \sim_{R'} \psem{p_4}{m^2} \sim_{R'} \psem{p_4}{m^3}$.

\fig{\ref{fig:analysis_ex}}c depicts the incorporation of the temporary equivalence relation~$R'$---the result of intra-instruction fault index coalescing---into equivalence relation~$R$ by the inter-instruction fault index coalescing analysis.
The result of intra-instruction fault index coalescing is applied across instructions only if the new fault indices do not conflict with other access points.
For instance, in \fig{\ref{fig:analysis_ex}}c, soft errors at data point~$v$ after program point~$p_2$ and before program point~$p_3$ affect all three reads of data point~$v$ at program points~$p_3$, $p_5$, and $p_6$. Hence, $\flowuse(p_2, v)=\{p_3, p_5, p_6\}$. But soft errors between program point~$p_2$ and $p_5$ or $p_6$ only affect data point~$v$ at program point~$p_5$ or $p_6$, therefore, $\flowuse(p_3, v)=\{p_5, p_6\}$.
Thus, $[\psem{p_3}{v^j}]_{R'}$, $[\psem{p_5}{v^j}]_{R'}$, and $[\psem{p_6}{v^j}]_{R'}$ are merged into $[\psem{p_2}{v^j}]_{R}$ only if $\psem{p_3}{v^j} \sim_{R'} \psem{p_5}{v^j} \sim_{R'} \psem{p_6}{v^j}$ for any $0 \le j < 4$. In \fig{\ref{fig:analysis_ex}}c, $[\psem{p_2}{v^2}]_{R}$ and $[\psem{p_2}{v^3}]_{R}$ are coalesced to $[\sem_0]_{R}$ after inter-instruction fault index coalescing, but $[\psem{p_2}{v^0}]_{R}$ and $[\psem{p_2}{v^1}]_{R}$ remain the same.

\section{Validation}\label{sec:validation}

The BEC analysis comprises two monotonic data-flow analyses, the bit-value analysis and the fault index coalescing analysis, and both are the maximal fixed-point~(MFP) assignment~\cite{Khedker:2009, Kam:1977}. In the bit-value analysis, each fault site meets \emph{all} definitions reachable to the fault site under the given CFG and overapproximates its bit-value. The fault index coalescing analysis determines two equivalence classes are equivalent only when it holds for \emph{all} use sites that may follow the fault site under the given CFG. The maximal fixed point guarantees the minimally acceptable soundness criterion of the quality of flow analysis~\cite{Wegman:const:prop, Graham:1975}.

We implemented the BEC analysis in LLVM~16.0.0 for the \makebox{RISC-V} architecture~\cite{RISCVISA}.
To validate the correctness of the implementation, we conducted exhaustive fault injection campaigns on execution traces of programs using an instrumented version of the SPIKE \makebox{RISC-V} ISA simulator~\cite{spike}. Each fault injection run induces a single soft error at a fault site, which is a particular bit of a register file at a specific clock cycle. For each fault site, individual runs are conducted to probe every bit of the register file exhaustively and one fault injection run is conducted per probed bit.
For each fault injection run, the location and the cycle of the fault to be injected are passed as command-line arguments to the ISA simulator. The ISA simulator executes the program for each fault injection run from the beginning, flips the to-be-probed bit when the specified execution cycle is reached, and resumes execution.

\begin{table}[t]
    \tabfontsize
    \centering
    \caption{Time and disk space requirements for \\the exhaustive fault injection campaign}
    \label{tab:fi}
    \begin{tabular}{|c|r|r|}
        \hline
        Benchmark & \multicolumn{1}{c|}{Time} & \multicolumn{1}{c|}{Disk space} \\
        \hline
        bitcount & \SI{0.5}{\hour}& \SI{1}{\giga\byte} \\
        AES & \SI{2}{\hour}& \SI{7}{\giga\byte} \\
        CRC32 & \SI{7}{\hour}& \SI{116}{\giga\byte} \\
        SHA & \SI{10}{\hour}& \SI{100}{\giga\byte} \\
        RSA & \SI{50}{\hour}& \SI{700}{\giga\byte} \\
        \hline
    \end{tabular}
\end{table}

Corrupted execution traces are generated for each fault injection run
and labeled based on their semantics. 
An execution trace comprises a sequence of executed instructions, side effects caused by the instructions executed such as memory accesses, and observable outcomes of the program.

The validation process highlights several merits of the BEC analysis. The
exhaustive fault injection campaign is a highly time- and disk-space consuming
process. Table~\ref{tab:fi} lists time and disk space required to conduct
fault injection campaigns on \emph{a single execution trace} per benchmark,
limited to benchmarks where the baseline campaign is tractable. Execution times
were obtained on a \SI{3.8}{\giga\hertz} AMD processor and execution traces are
generated and classified on the fly and only distinguishable traces are
archived to conserve disk space. Besides the time and space costs required for
fault injection campaigns, which can easily become infeasible even with small
programs, a fault injection campaign is required for each execution trace.
Considering that different initial machine states, e.g., due to program input, may
generate different execution traces for a program, it is impossible to test all
possible execution traces because the input space is generally
infinite~\cite{Preda:2020a,Preda:2020b}. In contrast, the BEC analysis needs to 
run only once per benchmark, at compile-time, and the analysis results hold for
all initial machine states and all possible execution traces. The BEC analysis
was tractable for all benchmarks, and no significant compile time overhead was
observed.

\begin{table}[t]
    \tabfontsize
    \centering
    \setlength\extrarowheight{3pt}
    \caption{Classification of Comparisons%
    }\label{tab:validation_categories}
    \begingroup
    \setlength{\tabcolsep}{0.6mm}
    \begin{tabular}{|@{~}c@{~}|c|}
    \hline
     \hspace{-0.6mm}\multirow{2}{*}{$\eqcl*{\psem{p}{v^i}}_R = \eqcl*{\psem{q}{u^j}}_R \wedge \ptrc{p}{v^i} = \ptrc{q}{u^i}$}    &  Sound,\\
                & precise \\[2pt]
     \hline
     \hspace{-0.6mm}\multirow{2}{*}{$\eqcl*{\psem{p}{v^i}}_R \neq \eqcl*{\psem{q}{u^j}}_R \wedge \ptrc{p}{v^i} = \ptrc{q}{u^j}$} &  Sound, \\
                     & imprecise \\[2pt]
     \hline
     \hspace{-0.6mm}$\eqcl*{\psem{p}{v^i}}_R = \eqcl*{\psem{q}{u^j}}_R \wedge \ptrc{p}{v^i} \neq \ptrc{q}{u^j}$ &  Unsound \\[2pt]
     \hline
    \end{tabular}
    \endgroup
\end{table}

\begin{table*}[!ht]
    \tabfontsize
    \centering
    \caption{Results of fault injection pruning by the proposed static analysis}
    \label{tab:fi_pruning}
    
    \begin{tabular}{|c|r|r|r|r|r|r|r|r|}
    \cline{2-9}
     \multicolumn{1}{c|}{}& \multicolumn{1}{c|}{bitcount} & \multicolumn{1}{c|}{dijkstra} & \multicolumn{1}{c|}{CRC32} & \multicolumn{1}{c|}{adpcm enc} & \multicolumn{1}{c|}{adpcm dec} & \multicolumn{1}{c|}{AES} & \multicolumn{1}{c|}{RSA} & \multicolumn{1}{c|}{SHA}\\
    \hline
    Live in values  &  \num{26272} & \num{230336} &  \num{245760} & \num{2819904} & \num{2003744} & \num{150112} & \num{1026304} & \num{421632} \\
    Live in bits    &  \num{20571} & \num{229409} &  \num{211176} & \num{2424874} & \num{1653714} & \num{105025} & \num{1025436} & \num{371294} \\
    \hline
    Masked bits     &   \num{2506} &     \num{70} &    \num{7368} &   \num{71000} &  \num{258000} &    \num{680} &     \num{434} &  \num{10660} \\    
    Inferrable bits &   \num{3195} &    \num{857} &   \num{27216} &  \num{324030} &   \num{92030} &  \num{44407} &     \num{434} &  \num{39678} \\
    \hline
    \hline
    \bf{Total FI runs pruned} & \bf{21.70\%} & \bf{0.40\%} & \bf{14.07\%} & \bf{14.01\%} & \bf{17.47\%} & \bf{30.04\%} & \bf{0.08\%} & \bf{11.94\%} \\
    \hline
    \end{tabular}
\end{table*}

To address the quality of the analysis, let us define $\ptrc{p}{v^i}$ to represent an execution trace generated from a fault injection run with a fault injected at fault site $(p, v^i)\in F$.
\tab{\ref{tab:validation_categories}} lists the three classifications of validation results where $(p, v^i), (q, u^j) \in F$.
The BEC analysis is considered sound and precise if it identifies all fault sites that generate identical execution traces when subjected to fault injection.
If the BEC analysis identifies some fault sites as not equivalent but the execution traces are identical, then the BEC analysis is sound but imprecise. This can occur in the presence of dynamic information which is not available at compile time, such as program inputs.   
We observed a negligible number of sound but imprecise cases, for instance, when analyzing global registers where no assumption is safe, such as \texttt{tp}, and \texttt{gp} in \makebox{RISC-V}.
If two fault injection runs are classified as identical but differ in their execution traces, the analysis is unsound. No such case was observed with the BEC analysis.

\section{Experimental Results}\label{sec:experimental_results}
The two use cases of the BEC analysis are evaluated using eight distinctive benchmarks from FISSC~\cite{FISSC:2016} and Mi\-Bench~\cite{mibench:2001}. 
We use integer benchmarks for the experiments, but the proposed approach is not limited to specific data types. For instance, x86~AVX registers~\cite{intel:arch:manual} provide bit-wise and/or operations on IEEE754 floats and can benefit from the proposed method in the same way as integer registers.

This section describes the details of the two use cases and their experimental results.

\subsection{Use Case 1: Fault Injection Campaign Pruning}\label{sec:usecase:fi_pruning}

\begin{table*}[!ht]
    \tabfontsize
    \centering
    \caption{Changes in the reliability against soft errors from bit-level vulnerability-aware instruction scheduling}
    \begin{tabular}{|c|r|r|r|r|r|r|r|r|r|}
    \cline{2-9}
     \multicolumn{1}{c|}{}& \multicolumn{1}{c|}{bitcount} & \multicolumn{1}{c|}{dijkstra} & \multicolumn{1}{c|}{CRC32} & \multicolumn{1}{c|}{adpcm enc} & \multicolumn{1}{c|}{adpcm dec} & \multicolumn{1}{c|}{AES} & \multicolumn{1}{c|}{RSA} & \multicolumn{1}{c|}{SHA}\\
    \hline
    Total fault space& \num{541696} & \num{27286528} &  \num{2922496} & \num{58426368} & \num{44085248} & \num{3180544} & \num{18295808} & \num{7483392} \\
    Best reliability &  \num{85018} &   \num{159966} &   \num{348384} & \num{28401348} & \num{19400720} & \num{1928214} &  \num{8650606} & \num{2559116} \\
    Worst reliability&  \num{94366} &   \num{166074} &   \num{394040} & \num{28530244} & \num{19538104} & \num{2007194} &  \num{8764640} & \num{2688188} \\
    \hline
    Worst/Best & \SI{111.00}{\percent} & \SI{103.82}{\percent} & \SI{113.11}{\percent} & \SI{100.45}{\percent} & \SI{100.71}{\percent} & \SI{104.10}{\percent} & \SI{101.32}{\percent} & \SI{105.04}{\percent}\\
    \hline
    \bf{+} & \bf{+11.00\%} & \bf{+3.82\%} & \bf{+13.11\%} & \bf{+0.45\%} & \bf{+0.71\%} & \bf{+4.10\%} & \bf{+1.32\%} & \bf{+5.04\%} \\
    \hline
    \end{tabular}
    \label{tab:vulnerability}
\end{table*}

The BEC analysis classifies the effect of soft errors across all fault sites of a program. Thus, the analysis result can be useful to prune fault injection runs of fault injection campaigns that are known to be masked or identical to the runs that have already been conducted.


\tab{\ref{tab:fi_pruning}} shows the number of fault sites that are identified as live and subject to fault injection runs without the BEC analysis~(Row~``Live in values'') and with the BEC analysis~(Row~``Live in bits'') on the eight benchmarks.
The numbers for ``Live in values'' are obtained based on the
inject-on-read\cite{Berrojo:2002,Schirmeier:2015,Sangchoolie:2015} analysis,
which employs value-level analysis to identify fault sites that are live and subject to fault injection runs.
Row~``Live in bits'' shows the numbers of live fault sites that
require fault injection runs when analyzed at the granularity of bits by the
BEC analysis. Row~``Total FI runs pruned'' depicts the percentages of fault
injection runs pruned by the BEC analysis~(rows~``Live in bits'' versus ``Live
in values''). Across the eight benchmarks, the BEC analysis pruned up to
\SI{30.04}{\percent} of fault injection runs, at an average of 
\SI{13.71}{\percent}.

Rows~``Masked bits'' and ``Inferrable bits'' in \tab{\ref{tab:fi_pruning}} show
the breakdown of fault sites pruned by the BEC analysis: Row~``Masked bits''
depicts the numbers of fault sites pruned because soft errors on those
fault sites are analyzed to be masked and dead, and Row~``Inferrable bits''
presents the numbers of fault sites pruned because the effects of the faults
are identical to other fault injection runs.

The degree of the effect of the BEC analysis on fault injection pruning varies due to benchmark characteristics.
Bit values are rarely known in AES at compile-time. Yet, AES frequently uses \texttt{xor} operations to encrypt or decrypt secret keys, which is particularly effective for the BEC analysis. With \texttt{xor} operations, fault indices coalesce unconditionally across instructions.
These characteristics of AES resulted in the highest pruning rate across
the set of benchmarks, \SI{30.04}{\percent}.

BEC achieved a solid \SI{17.47}{\percent} pruning rate for the ADPCM
decoder~(adpcm dec), but for a different reason. 
The decoding process involves an abundant number of bit operations, more importantly, with constant values.
This enhances the quality of bit-value
analysis, leading to more precise analysis results. The ADPCM encoder and
decoder perform internal operations on 4-bit values but clamp to 1-bit or 2-bit
values to output. Such characteristics of the benchmarks foster the BEC
analysis to identify more fault sites to be masked.


RSA is an adversary case for the BEC analysis because the majority of its
operations are arithmetic and thus challenging for bit-value analysis, with
noticable impact on the performance of the fault index coalescing analysis.


%
%
\subsection{Use Case 2: Bit-level Vulnerability-aware Instruction Scheduling}\label{sec:usecase:vul_anal}
Vulnerability-aware instruction scheduling highlights the merit of static analysis. 
The BEC analysis identifies which fault sites are masked and insensitive to soft errors, and such information can be employed with instruction scheduling to enhance the reliability of programs against soft errors.
Instruction scheduling in LLVM is based on list scheduling~\cite{SCHUTTEN:1996}. First, it maintains a dependency graph among instructions and registers accessed per basic block to examine which instructions in a basic block are ready to be scheduled. If multiple instructions are ready for instruction scheduling, LLVM uses heuristics to choose the one with the highest priority. These heuristics include register pressure, instruction latency, pipeline stalls, etc. The number of fault sites susceptible to soft errors analyzed by the BEC analysis is used as a novel criterion to select the most appropriate instruction to schedule when there is more than one candidate.

\begin{algorithm}
\caption{Instruction Scheduling for Reliability}\label{algo:instr_sched}
\SetKwInput{KwInput}{Input}
\SetKwInput{KwOutput}{Output}
\DontPrintSemicolon

\KwInput{a set of instructions $P$ with data dependencies}
\KwOutput{a sequence of instructions $S$}
Initialize an empty schedule $S$.

\While{$P \neq \emptyset$} {
  Choose a set of ready instructions $R$ in $P$ with no unsatisfied data dependencies\;  
  Choose an instruction $p$ in $R$ which kills the most fault sites in bits\;  
  Add the instruction $p$ to $S$\;
  Remove the instruction $p$ from $P$\;
}
\end{algorithm}

Algorithm~\ref{algo:instr_sched} describes how instruction scheduling is
conducted with the BEC analysis as the selection criterion. For each scheduling
region, i.e., a basic block, a data dependency graph is created with
instructions. The instruction scheduler iterates over the set of instructions
until all the instructions are scheduled into a list of instructions. When
multiple instructions satisfy the data dependency requirement and are ready to
be scheduled, the result of the BEC analysis is used as a selection criterion,
and the instruction that reduces the most unmasked fault sites is prioritized
for scheduling.

\tab{\ref{tab:vulnerability}} shows the experimental results when the BEC analysis is used as the new selection criterion for instruction scheduling. Row~``Total fault space'' in \tab{\ref{tab:vulnerability}} indicates the total number of fault sites per execution trace of benchmarks, and Row~``Best reliability'' in \tab{\ref{tab:vulnerability}} shows the numbers of fault sites susceptible to soft errors when the instruction scheduling criterion is set to maximize the number of masked fault sites. Row~``Worst reliability'' shows the numbers of fault sites susceptible to soft errors when the selection criterion is the opposite.
Row~``Worst/Best'' in \tab{\ref{tab:vulnerability}} shows the maximum possible reliability improvement against soft errors with the proposed instruction scheduling technique.


The benchmarks that improved noticably from vulnerability-aware
instruction scheduling were CRC32 and bitcount, showing \SI{13.11}{\percent}
and \SI{11.00}{\percent} reduction in vulnerable fault sites, respectively.
Both CRC32 and bitcount have abundant numbers of masked bits, facilitating a high
improvement in reliability against soft errors after vulnerability-aware
instruction scheduling. 

Both the ADPCM encoder and decoder contain large numbers of masked bits, but
instructions are rather tightly ordered compared to other benchmarks. It
restricted scheduling flexibility irrespective of the respective 
scheduling policy.

AES does not contain many bits masked in live registers. However, the registers
stay live relatively long for infrequent read accesses. The effect of masked
bits are exacerbated in such long-lived registers even if the masked bits are
few. Shortening the live ranges of registers with more live fault sites
contributed the most to the improvement of reliability in the case of AES. 

RSA lacks masked or inferable bits to be exploited by the new scheduling
criterion and it is a highly sequential program with frequent memory accesses.
Thus, there was little room for improvement by varying instruction scheduling
criteria, yet the proposed instruction scheduling technique achieved a
\SI{0.08}{\percent} improvement. 



Bit-level vulnerability-aware instruction scheduling
enhanced by the BEC analysis increased the reliability of programs by up to \SI{13.11}{\percent},
and \SI{4.94}{\percent} on average.
Despite the heuristic nature of instruction scheduling, no degradation of the 
reliability
against soft errors was observed among the benchmarks that we have evaluated.
It is worth noting that instruction scheduling does not affect the number
of instructions executed per program nor the raw number of fault injection runs
required for an exhaustive fault injection campaign, because it does not change
the number of data accesses. 


%
%
%
\section{Related Work}\label{sec:related_work}

%
%

%
%
\subsection{Static vs.\ Dynamic Analysis for Reliability}
Several dynamic approaches have been proposed to analyze the propagation of faults at bit-level in hardware components~\cite{Dietrich:2018,Berrojo:2002} and in execution traces of programs~\cite{Pusz:2021}. Dynamic analysis under-approximates program semantics. In contrast, static analysis, which is the proposed method, computes an over-approximation of program semantics. Thus, a single static analysis run covers all possible inputs or execution traces of a program, while dynamic analysis must be performed for every possible input or trace of a program. The essential limitation with dynamic approaches is therefore that it is impossible to test all possible inputs because the input space is generally infinite~\cite{Preda:2020a,Preda:2020b}. 

Another merit of static analysis is that it can contribute to improving programs fundamentally by optimizing or transforming the program at compile-time based on the analysis results. As a proof of concept, we have conveyed the result of the proposed analysis method to the instruction scheduler of LLVM to enhance the reliability of programs against soft errors.

%
%
\subsection{Fault Injection Pruning}
We have demonstrated the effectiveness of the BEC analysis on fault injection pruning as the first use case.
One efficient and widely used fault injection pruning strategy is to conduct fault injection runs just before the value is read. This method was introduced by Smith~et~al.~\cite{Smith:1995} as a way to determine equivalent fault classes for permanent and transient faults in values. It was further developed to Inject-on-read~\cite{Berrojo:2002,Schirmeier:2015,Sangchoolie:2015,Ko:2022}. 
Unlike BEC, these methods operate at the granularity of values and hence lack
optimization opportunities at bit-level.



%
%
\subsection{Instruction Scheduling for Reliability}
As the second use case of the proposed bit-level analysis, we used the analysis results as the criteria for instruction scheduling in LLVM. With the results of the proposed bit-level analysis, instruction scheduling can be determined in a way to maximize the number of fault sites insensitive to soft errors. Rehman~et~al.~\cite{Rehman:2012} proposed reliability-aware instruction scheduling strategies that determine reliability-critical instructions by looking ahead in the instruction sequences.
Xu~et~al.~\cite{Xu:2011} proposed an instruction scheduling strategy that reduces the overall length of the live intervals of registers. These approaches are based on value-level analysis for the criticality or longevity of registers. The proposed bit-level analysis can be readily combined with the previous approaches to further enhance the effectiveness.
Instruction scheduling augmented by the BEC analysis enhanced the reliablity of programs against soft errors comparable to the improvements achieved by established methods in the field~\cite{Yan:2005, Xu:2009,Xu:2011}.







\section{Conclusion}\label{sec:conclusion}
We have presented BEC, a static bit-level analysis that enhances the reliability of programs against soft errors, and two of its use cases. The work has been implemented within LLVM~16.0.0 for the \makebox{RISC-V} architecture and validated on an instrumented version of the SPIKE \makebox{RISC-V} ISA simulator using eight benchmarks with distinctive characteristics. 
The proposed bit-level analysis pruned up to \SI{30.04}{\percent}
of exhaustive fault injection campaigns (\SI{13.71}{\percent} on average), without loss
of accuracy. Program vulnerability reduced by up to \SI{13.11}{\percent} (\SI{4.94}{\percent} on average) through bit-level vulnerability-aware instruction scheduling. 

The BEC analysis has been applied to software, but we anticipate this work to be easily extended for hardware testing to reduce production costs. For instance, it can be adopted as a pass of hardware synthesis tools to shorten the chip testing process analytically and systematically.

\section{Data-Availability Statement}
The code that supports the findings of this study has been open-sourced on GitHub~\cite{BEC:github}
and archived on Zenodo~\cite{BEC:zenodo}.



\section*{Acknowledgment}
We thank the anonymous reviewers for their insightful comments and suggestions.
This work was partly supported by
the Institute of Information \& Communications
Technology Planning \& Evaluation~(IITP) grant funded by the Ministry of
Science and ICT~(MSIT)~(No.~2021-0-00853), and by the BK21 FOUR of the Department of
Computer Science and Engineering, Yonsei University, funded by the National
Research Foundation of Korea~(NRF).


\bibliographystyle{IEEEtranDOI} 
\bibliography{papers}


\end{document}